%% file: cpt_f.tex
\documentclass{ws-ijmpa}
\usepackage{epsf,graphics,epsfig,graphicx,feynarts}
\begin{document}

\def\cM{{\cal M}} 
\def\cO{{\cal O}}
\def\cK{{\cal K}}
\def\cS{{\cal S}}
\newcommand{\mh}{m_h}
\newcommand{\mw}{m_W}
\newcommand{\mz}{m_Z}
\newcommand{\mt}{m_t}
\newcommand{\mb}{m_b}
\def\lsim{\mathrel{\raise.3ex\hbox{$<$\kern-.75em\lower1ex\hbox{$\sim$}}}}
\def\gsim{\mathrel{\raise.3ex\hbox{$>$\kern-.75em\lower1ex\hbox{$\sim$}}}}
\def\ga{\mathrel{\raise.3ex\hbox{$>$\kern-.75em\lower1ex\hbox{$\sim$}}}}
\def\la{\mathrel{\raise.3ex\hbox{$<$\kern-.75em\lower1ex\hbox{$\sim$}}}}

\newcommand{\non}{\nonumber}
%============================================
%\vspace{3.cm}
%\noindent
%\vspace{3.cm}
%\preprint{LPHEA-06-03}
\markboth{A. Arhrib, R. Benbrik, M. Chabab, W.T. Chang and T.-C. Yuan}
{CP violation in 
Charged Higgs Bosons decays $H^\pm \to W^\pm (\gamma ,Z)$ 
in MSSM}

\catchline{}{}{}{}{}

\title{CP violation in 
Charged Higgs Bosons decays $H^\pm \to W^\pm (\gamma ,Z)$ 
in the Minimal Supersymmetric Standard Model (MSSM)}

\author{ABDESSLAM ARHRIB}
\address{D\'epartement de Math\'ematiques, Facult\'e des Sciences et 
Techniques, B.P 416 Tangier, Morocco\\
and National Central University,
Department of Physics,
Chung-li, Taiwan 32054, R.O.C.\\
aarhrib@ictp.it}
\author{RACHID BENBRIK}
\address{Chung Yuan Christian University,
Department of Physics,
Chung-Li, Taiwan 320, R.O.C.\\
LPHEA, D\'epartement de Physique,Facult\'e des Sciences-Semlalia B.P
2390 Marrakech, Morocco\\
rbenbrik@phys.cycu.edu.tw\\
}
\author{MOHAMED CHABAB}
\address{LPHEA, D\'epartement de Physique,
             Facult\'e des Sciences-Semlalia\\
             B.P 2390 Marrakech, Morocco\\
mchabab@ucam.ac.ma
}
\author{WEI TING CHANG AND TZU-CHIANG YUAN}
\address{Department of Physics, National Tsing Hua University,
Hsinchu, Taiwan 300, R.O.C.
}
\maketitle

\begin{abstract}
One loop mediated charged Higgs bosons decays
$H^\pm\to W^\pm V$, $V= Z, \gamma$ are studied in the Minimal 
Supersymmetric Standard Model (MSSM) with and without CP 
violating phases. We evaluate the MSSM contributions to these processes 
taking into account $B\to X_s\gamma$ constraint as well as experimental 
constraints on the MSSM parameters. In the MSSM, we found that 
in the intermediate range of $\tan\beta \la 10$ and for large 
 $A_t$ and large $\mu$, where the lightest top squark
becomes very light and hence non-decoupled, 
the branching ratio of $H^\pm \to W^{\pm} Z$ can be of the order 
$10^{-3}$ while the branching ratio of $H^\pm \to W^{\pm} \gamma$ 
is of the order $10^{-5}$.  We found also that the CP violating 
phases of soft SUSY parameters can modify the branching 
ratio by about one order of magnitude. We also show that
MSSM with CP violating phases lead to CP-violating asymmetry in 
the decays $H^+ \to W^+V$ and $H^- \to W^-V$. 
Such CP asymmetry can be rather large and can 
reach  80\% in some region of parameter space.

\end{abstract}

\section{Introduction}
\label{sec:intro}
Supersymmetric (SUSY) theories, in particular the Minimal Supersymmetric
Standard Model (MSSM), are currently considered as the 
most theoretically well motivated extensions of the Standard Model. 
Recently, phenomenology of the MSSM with complex SUSY parameters
has received growing attention \cite{MSSMCP}.
Such complex phases provide new sources of CP violation 
which may explain electroweak baryogenesis scenarios \cite{Baryogen}, 
and CP violating phenomena in $K$ and $B$ decays \cite{baek}.
It has been shown in \cite{dugan} 
that by assuming universality of the gaugino masses at
a high energy scale, the effects of complex soft SUSY 
parameters in the MSSM can be parameterized 
by two independent CP violating phases: the phase of the 
Higgsino mass term $\mu$ (Arg($\mu$)) and the phase 
of the trilinear scalar coupling parameters $A=A_f$ (Arg($A_f$)) 
of the sfermions $\widetilde{f}$. 
The presence of large SUSY phases can give contributions to 
electric dipole moments of the electron and neutron (EDM) 
which exceed the experimental upper bounds. In a variety of SUSY 
models such phases turn out to be severely restricted 
by such constraints,  {\it i.e.} ${\rm Arg}(\mu) < {\cal }(10^{-2})$ for a SUSY
mass scale of the order of few hundred GeV \cite{nath}.
However, the possibility of having large CP violating phases 
can still be consistent with experimental data in any of the 
following three scenarios: i) Effective SUSY models \cite{nath},
ii) Cancellation mechanism \cite{cancell} and iii) Non-universality of 
trilinear couplings $A_f$ \cite{trilin}.
In the MSSM, after electroweak symmetry
breaking we are left with 5 physical Higgs particles 
(2 charged Higgs $H^\pm$, 2 CP-even $H^0$, $h^0$
and one CP--odd $A^0$). In this study, our concerns is
 the charged Higgs decays $H^\pm\to W^\pm V$, $V= Z, \gamma$ 
with and without CP violating phases. Those channels have
a very clear signature and might emerge easily at future colliders.
For instance, if $H^\pm \to W^\pm Z$ is enhanced enough, this decay 
may lead to three leptons final state if both $W$ and $Z$ decay leptonically
and that would be the corresponding golden mode for charged Higgs boson.
Charged Higgs phenomenology has been extensively studied in the 
literature. It has been shown  that 
SUSY one-loop contributions can lead to decay 
rates asymmetry of $H^\pm \to t\bar{b}, b\bar{t}$, 
$H^\pm \to \widetilde{u}_i\widetilde{d}_j^*,\widetilde{u}_i^*\widetilde{d}_j$ 
and $H^\pm \to \widetilde{\chi}_i^+\widetilde{\chi}_j^0,
\widetilde{\chi}_i^-\widetilde{\chi}_j^0 $ \cite{chris} . 
Similar study has been done for the single charged Higgs 
production at hadron collider \cite{cp} and shown that SUSY 
CP phases can lead to CP asymmetry.

In this paper, we will discuss both the branching ratios of 
 $H^\pm\to W^\pm V$ with $V= Z, \gamma$ \cite{abc}
as well as the CP-violating asymmetry in the decay rates $H^+ \to W^+V$ 
and $H^- \to W^- V$ that emerge from the presence of CP violating 
phases in MSSM.

The paper is organized as follows. In section II, we 
describe our calculations and the one-loop renormalization scheme
we will use for $H^\pm\to W^\pm V$ and present the source of 
CP violation in $H^\pm\to W^\pm V$. In Section III, we present our numerical
results and discussions, and section IV contains our conclusions.

\section{CP violation in Charged Higgs decay: 
$H^\pm \to W^\pm V$}
\subsection{One loop amplitude $H^\pm \to W^\pm V$}
In MSSM, at tree level, the coupling $H^\pm W^\pm \gamma$
is absent because of electromagnetic gauge invariance $U(1)_{\rm em}$,
while the absence of $H^\pm W^\pm Z $ is due to the 
isospin symmetry of the kinetic Lagrangian of the Higgs doublet fields.
We emphasize here that it is possible to 
construct models with an even larger scalar sector than 
2 Higgs doublets. 
One of the most popular being the Higgs Triplet
Model (HTM) \cite{triplet}. 
A noteworthy difference between MSSM and HTM is that 
the HTM contains a tree level $ZW^\pm H^\mp$ coupling.
Therefore, in MSSM, decays modes like $H^\pm \to W^\pm \gamma$, 
$H^\pm \to W^\pm Z $ are mediated at one loop level 
\cite{pom,micapey,ray,kanemur,toscano}. 
Although these decays are rare processes, 
loop or/and threshold effects can give rise to substantial enhancement.
Moreover, once worked out, any experimental deviation from the results 
within such a model should bring some fruitful information on the new 
physics and also allow to distinguish between models. 
Let us give the general structure of the 
one loop amplitude for $H^\pm\to W^\pm V$  
and discuss the renormalization scheme introduced
to deal with tadpoles and vector boson-scalar boson mixing.

The amplitude ${\cal M}$ for a scalar decaying to two
gauges bosons $V_1$ and $V_2$ can be written as
\begin{eqnarray}
{\cal M}=\frac{g^3\epsilon_{V_1}^{\mu *}\epsilon_{V_2}^{\nu *}}
{16\pi^2 m_W}{\cal M}_{\mu\nu}
\end{eqnarray}
where $\epsilon_{V_i}$ are the polarization vectors of the $V_i$.
According to Lorentz invariance, the general structure of 
the one loop amplitude ${\cal M}_{\mu\nu}$ of 
$S\to V_1^{\mu} V_2^{\nu}$ decay is
%if CP is conserved, 
%
\begin{eqnarray}
{\cal M}_{\mu\nu}(S\to W^\mu V^\nu) =
{\cal F}_1 g_{\mu\nu} + {\cal F}_2 p_{1\mu} p_{2\nu} + 
{\cal F}_3 
i\epsilon_{\mu\nu\rho\sigma}p_{1}^{\rho}p_{2}^{\sigma}\label{lor}
\end{eqnarray}
where $p_{1,2}$ are the momentum of $V_{1}$, $V_{2}$ vector bosons, 
${\cal F}_{1,2,3}$ are form factors, and 
$\epsilon_{\mu\nu\rho\sigma}$ is the totally antisymmetric tensor.
The form factor ${\cal F}_1$ has dimension 2 while the other are 
dimensionless.
Therefore, it is expected that in case of $H^\pm \to W^\pm Z$ decay, 
${\cal F}_1$ will grow quadratically with internal top quark mass
while ${\cal F}_{2,3}$ will have only logarithmic dependence
\cite{micapey}.\footnote{As it has been shown in Ref.~[\refcite{micapey}], 
the top quark contribution does not decouple while squarks 
contributions does.}
Quite contrary, for $H^\pm\to W^\pm \gamma$ decay, the
electromagnetic gauge invariance implies that ${\cal F}_{1}$ and 
${\cal F}_{2}$ are related by
${\cal F}_1=$$1/2(m_W^2-m_{H\pm}^2){\cal F}_2$ \cite{micapey}.
This means that only  ${\cal F}_2$ and ${\cal F}_3$ will 
contribute to the decay $H^\pm \to W^\pm \gamma$ 
and then the amplitude of $H^\pm \to W^\pm \gamma$
will not grow quadratically with internal fermionic masses.
In case of $H^\pm \to W^\pm Z$, there is no such constraint on the form factors.
In fact the CP asymmetry in this process $H^\pm \to W^\pm Z$
is very large compared to the CP asymmetry 
in $H^+\to t\bar{b}$ \cite{chris}. The reason is that $H^\pm \to W^\pm Z$
 is loop dominated, while $H^+\to t\bar{b}$ is  dominated
 by the tree level contribution which is CP conserving.

%--------------------------------------------------
\begin{figure}[t!]
\begin{center}
\vspace{-3.3cm}
\input{vertz_MSSM.tex}
\vspace{-2.2cm}
\caption{Generic contributions to $H^\pm \to W^{\pm}V$}
\label{hwz}
\end{center}
\end{figure}
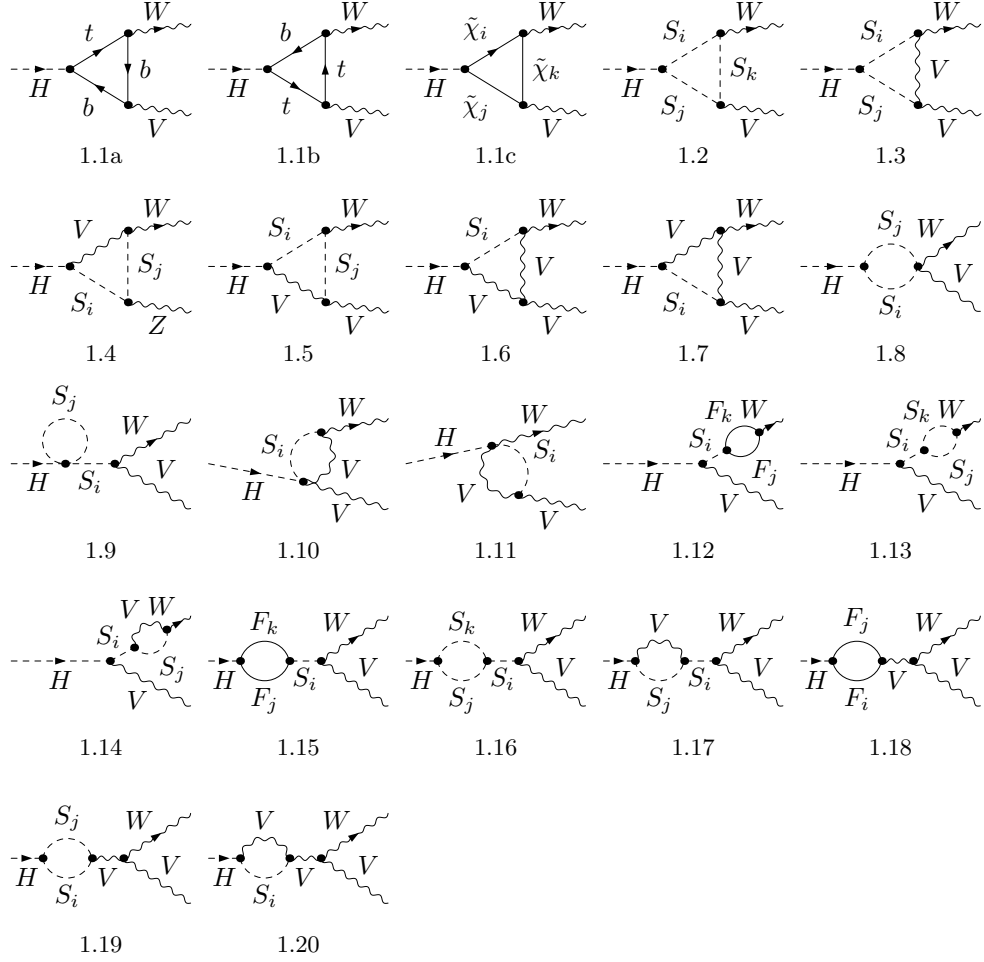
%-------------------------------------------------------
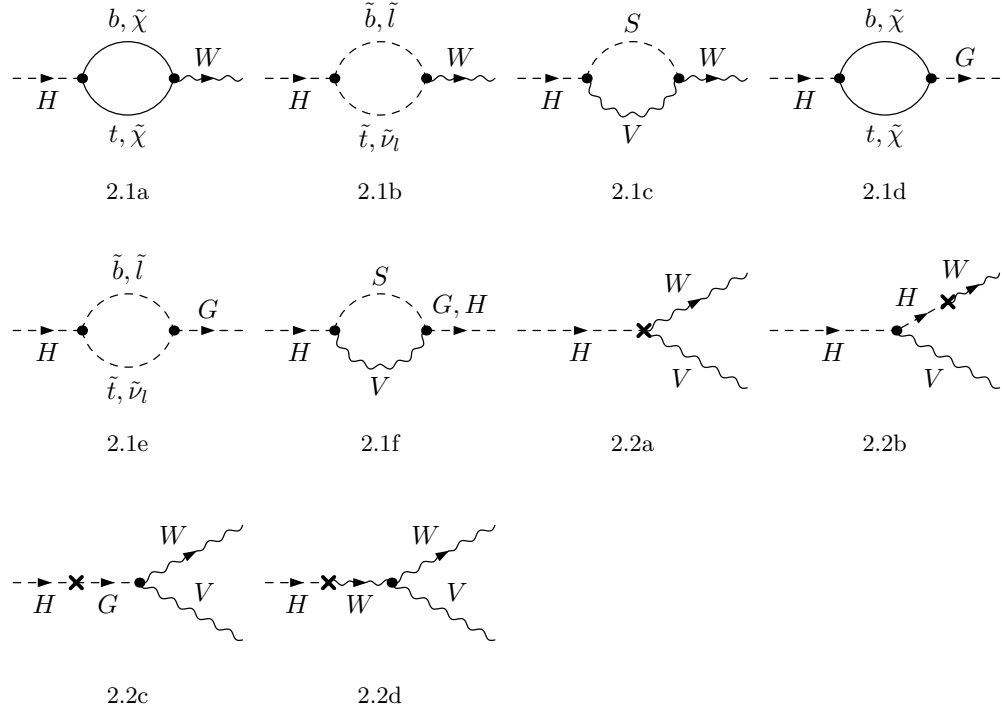
\begin{figure}[t!]
\begin{center}
\vspace{-2.cm}
\input{self.tex}
\vspace{-6.6cm}
\caption{Generic contributions to $H^\pm \to W^{\pm}$ and
$H^\pm \to G^{\pm}$ mixing as well as counter-terms needed.}
\label{hwmix}
\end{center}
\end{figure}
%-------------------------------------------------------

The one-loop amplitude for 
$H^\pm \to  W^\pm V$ is evaluated in the 't Hooft-Feynman gauge using
dimensional regularization. 
The  typical Feynman diagrams that contribute to  $H^\pm \to  W^\pm V$
are depicted in Fig.~1. Those diagrams contain vertex diagrams
(Fig.~1.1 $\to$ 1.11), $W^\pm$-$H^\pm$ mixing (Fig.~1.12 $\to$ 1.14),
$H^\pm$-$G^\pm$ mixing (Fig.~1.15 $\to$ 1.17) and
$H^\pm$-$W^\pm$ mixing (Fig.~1.18 $\to$ 1.20).

Note that the mixing $H^\pm$--$W^\pm$ (Fig~.1.12, 1.13, 1.14) vanishes for 
an on-shell transverse W gauge boson.  
There is no contribution from the $W^\pm$--$G^\mp$ 
mixing because $\gamma G^\pm H^\mp$ and $ZG^\pm H^\mp$ vertices 
are absent at the tree level. 
All the Feynman diagrams have been generated and computed using 
FeynArts and FormCalc \cite{seep} packages.
We also used  the fortran FF--package \cite{ff} in the numerical 
analysis.

Although the amplitude for our process is absent at the tree level,
complications like tadpole contributions and 
vector boson--scalar boson mixing require a careful treatment of 
renormalization. We adopt, hereafter, the on-shell renormalization 
scheme of Ref.~[\refcite{dabelstein}], for the Higgs sector. 
In this scheme,  field renormalization is performed in the 
manifest-symmetric version of the Lagrangian. 
A field renormalization constant is assigned to each 
Higgs doublet $\Phi_{1,2}$. 
We follow closely the same approach adopted in Refs.~[\refcite{abc,achm}] for 
$W^\pm$-$H^\mp$ and $H^\pm$-$G^\mp$ mixing as well as renormalization
of $\gamma W^\pm H^\mp$ and Z$W^\pm H^\mp$. 
We will use the following on shell renormalization conditions \cite{abc,achm}:
\begin{itemize}
\item  The renormalized tadpoles, {\it i.e.} the sum of
      tadpole diagrams $T_{h,H}$ and tadpole counter-terms
      $\delta_{h,H}$ vanish:   
\[  T_{h} +\delta t_h=0, \quad T_H +\delta t_H=0 \, . \]
These conditions guarantee that the vevs $v_{1,2}$ appearing in the renormalized 
Lagrangian ${\cal L_R}$ are located at the minimum of the one-loop potential.
\item The real part of the renormalized 
non-diagonal self-energy $\hat{\Sigma}_{H^\pm W^\pm}(k^2)$ 
vanishes for an on-shell charged Higgs boson. 
This renormalization condition determines the 
counter term for 
$\delta [A_\nu W_\mu^\pm H^\mp]$ 
and $\delta [Z_\nu W_\mu^\pm H^\mp]$.
\end{itemize} 
Using the Slavnov--Taylor identity, which relates $H^\pm W^\pm $  and 
$H^\pm G^\pm$ self energies,
the last renormalization condition about $H^\pm W^\pm $ mixing 
is sufficient to discard the real part of 
the $H^\pm$--$G^\pm$ mixing contribution
as well.

\subsection{CP violation in  $H^\pm \to W^\pm V$}
The CP violating phases of MSSM can lead to 
CP-violating asymmetry in the decay rates $H^+ \to W^+V$ and 
$H^- \to W^-V$. In what follow, we will study 
the following CP-violating asymmetry 
\begin{eqnarray}
{\delta}^{CP}=\frac{\Gamma(H^+\to W^+V)- 
\Gamma(H^-\to W^-V) }{\Gamma(H^+\to W^+V)+\Gamma(H^-\to W^-V)}\qquad , \qquad
V=\gamma , Z\label{cpa}
\end{eqnarray}
and show that the MSSM with complex phases together 
with the absorptive part emerging from some loop integrals
can lead to non-vanishing ${\delta}^{CP}$.
Generically, in the case of $H^\pm\to W^\pm\gamma$, 
the form factors ${\cal F}_{1,3}$
take the following form
\begin{eqnarray}
& &{\cal F}_1 (H^-\to W^-\gamma) \propto 
\sum_{i,j,k} a_{1,ijk} {\rm{PV}}_{1,ijk} +
a_{2,ijk} {\rm{PV}}_{2,ijk}\nonumber\\
& &{\cal F}_3(H^-\to W^-\gamma) \propto \sum_{i,j,k} b_{1,ijk} 
\widetilde{\rm{PV}}_{1,ijk} +
b_{2,ijk}\widetilde{\rm{PV}}_{2,ijk}\nonumber\\
& & {\cal F}_1 (H^+\to W^+\gamma) \propto 
\sum_{i,j,k} a_{1,ijk}^\star {\rm{PV}}_{1,ijk} +
a_{2,ijk}^\star {\rm{PV}}_{2,ijk}\nonumber\\
& & {\cal F}_3(H^+\to W^+\gamma) \propto \sum_{i,j,k} b_{1,ijk}^\star 
\widetilde{\rm{PV}}_{1,ijk} +
b_{2,ijk}^\star \widetilde{\rm{PV}}_{2,ijk}\nonumber
\end{eqnarray}
where $(a,b)_{\alpha,ijk}$ ($\alpha=1,2$ and $i,j$ and $k$ label the particles inside the loop) are combination of MSSM couplings, 
namely charged Higgs couplings to charginos-neutralinos and 
 charged Higgs couplings to squarks. ${\rm{PV}}_{\alpha,ijk}$ and 
$\widetilde{\rm{PV}}_{\alpha,ijk}$ are Passarino-Veltman
functions. 
The CP asymmetry (\ref{cpa}) takes the following form
\begin{eqnarray}
{\delta}^{CP} &\propto & (|{\cal F}_1(H^-)|^2 - 
|{\cal F}_3(H^-)|^2 - |{\cal F}_1(H^+ )|^2 + 
|{\cal F}_3(H^+ )|^2)/{\mathrm all} \nonumber\\
 & \propto &\sum_{i,j,k}\sum_{\alpha\beta} 
\Im m (a_{\alpha, ijk}a_{\beta, ijk}^\star) \Im m({\rm{PV}}_{\alpha ,ijk}
{\rm{PV}}_{\beta ,ijk}^\star) + 
( \alpha \longleftrightarrow \beta , 
{\rm{PV}} \longleftrightarrow \widetilde{\rm{PV}})
\nonumber
\label{cpaa}
\end{eqnarray}
It is clear from the above equation that in order to have 
a non-vanishing ${\delta}^{CP}$ we need both complex couplings
$a_{\alpha,ijk}$ as well an absorptive part 
from the Passarino-Veltman functions. For example,
diagrams Fig.1.1c and Fig.1.18 (resp. Fig.1.2 and Fig.1.19)
will develop some absorptive parts if the decay
channels $H^\pm \to \widetilde{\chi}_i^\pm  \widetilde{\chi}_j^0$
(resp. $H^\pm \to \widetilde{q}_i^\pm  \widetilde{q'}_j^*$) is open,
{\it i.e.} $m_{H\pm} > m_{\widetilde{\chi}_i} m_{\widetilde{\chi}_j}$ 
(resp. $m_{H\pm} > m_{\widetilde{q}_i} m_{\widetilde{q'}_j}$). 

\section{Numerics and discussions}
In our numerical evaluations, we use the following experimental 
input quantities~\cite{pdg4}: $\alpha^{-1}=129$, $m_Z$, $m_W$, $m_t$,
$m_b=$ 91.1875, 80.45, 174.3, 4.7  GeV. In the MSSM,
we specify the free parameters that will be used as follow: 
$i)$ The MSSM Higgs sector is  parameterized by the CP-odd mass 
$m_{A^0}$ and $\tan\beta$, taking into account 
one-loop radiative corrections from [\refcite{okada}], 
and we assume $\tan\beta \ga 3$.
$ii)$ The chargino--neutralino sector can be 
parameterized by the  gaugino-mass  terms
$M_1$, $M_2$, and the Higgsino-mass term $\mu$. For simplification,  
GUT relation $M_1\approx M_2/2$ is assumed. 
$iii)$ Sfermions are characterized 
by a common soft-breaking sfermion mass 
$M_{SUSY} \equiv \widetilde{M}_L=\widetilde{M}_R$, 
$\mu$ the parameter and the soft trilinear couplings 
for third generation scalar fermions $A_{t,b,\tau}$. For 
simplicity,  we will take $A_t=A_b=A_\tau$. 

When varying the MSSM parameters, 
we take into account also the following constraints:
$i)$  The extra contributions to the $\delta\rho$ parameter from the 
Higgs scalars should not exceed the current limits from precision 
measurements \cite{pdg4}: $|\delta\rho|\la 0.003$.
$ii)$ $b \to s \gamma$ constraint. The present world average for 
inclusive $b\to s \gamma$ rate is \cite{pdg4}
${\cal B}(B \rightarrow X_s \gamma) = (3.3 \pm 0.4) \times 10^{-4}$. 
We keep the $B \rightarrow X_s
\gamma$ branching ratio in the 3$\sigma$ range of (2.1--4.5)$\times
10^{-4}$.  The SM part  of 
 $B \rightarrow X_s \gamma$ is calculated up to NLO using the 
expression given in [\refcite{NK}]. While for the MSSM part, the Wilson coefficients
$C_7$ and $C_8$ are included at LO in the framework of MSSM with 
CKM as the only source of flavor violation and are taken from 
[\refcite{c7c8}]. $iii)$ We will assume that all SUSY particles sfermions
and charginos are heavier than about 100 GeV; for the light CP even 
Higgs we assume $m_{h^0}\ga 98$ GeV and $\tan\beta\ga 3$ \cite{lepp}.
For charged Higgs boson we assume that $m_{H\pm}\geq 78$ GeV from
LEP experiments \cite{LEP1}.
As the experimental bound on $m_{h^0}$ is concerned, care has to be taken. 
Since we are using only one-loop approximation for the 
Higgs spectrum, and as is well known,  higher order corrections \cite{sven}
may reduce the light CP-even Higgs mass in some cases. It may be possible
that some parameter space region, shown in this analysis, 
which survives the experimental limit 
$m_{h^0}\ga 98$ GeV with one loop calculation may disappear, once the higher
order correction to the Higgs spectrum are included.

The total width of the charged Higgs is computed at tree level from
[\refcite{abdel2}] without any QCD improvement for its fermionic decays 
$H^\pm\to \bar{f}f'$. The SUSY channels
like $H^+\to \widetilde{f}_i \widetilde{f}_j'$ and 
$H^+\to \widetilde{\chi}_i^0\widetilde{\chi}_j^+$ are
included when kinematically allowed.
%%%%%%%%%%%%%%%%%%%%%%%%%%%%%%%%
\begin{figure}
\begin{center}
\begin{tabular}{cc}
\resizebox{78mm}{!}{\includegraphics{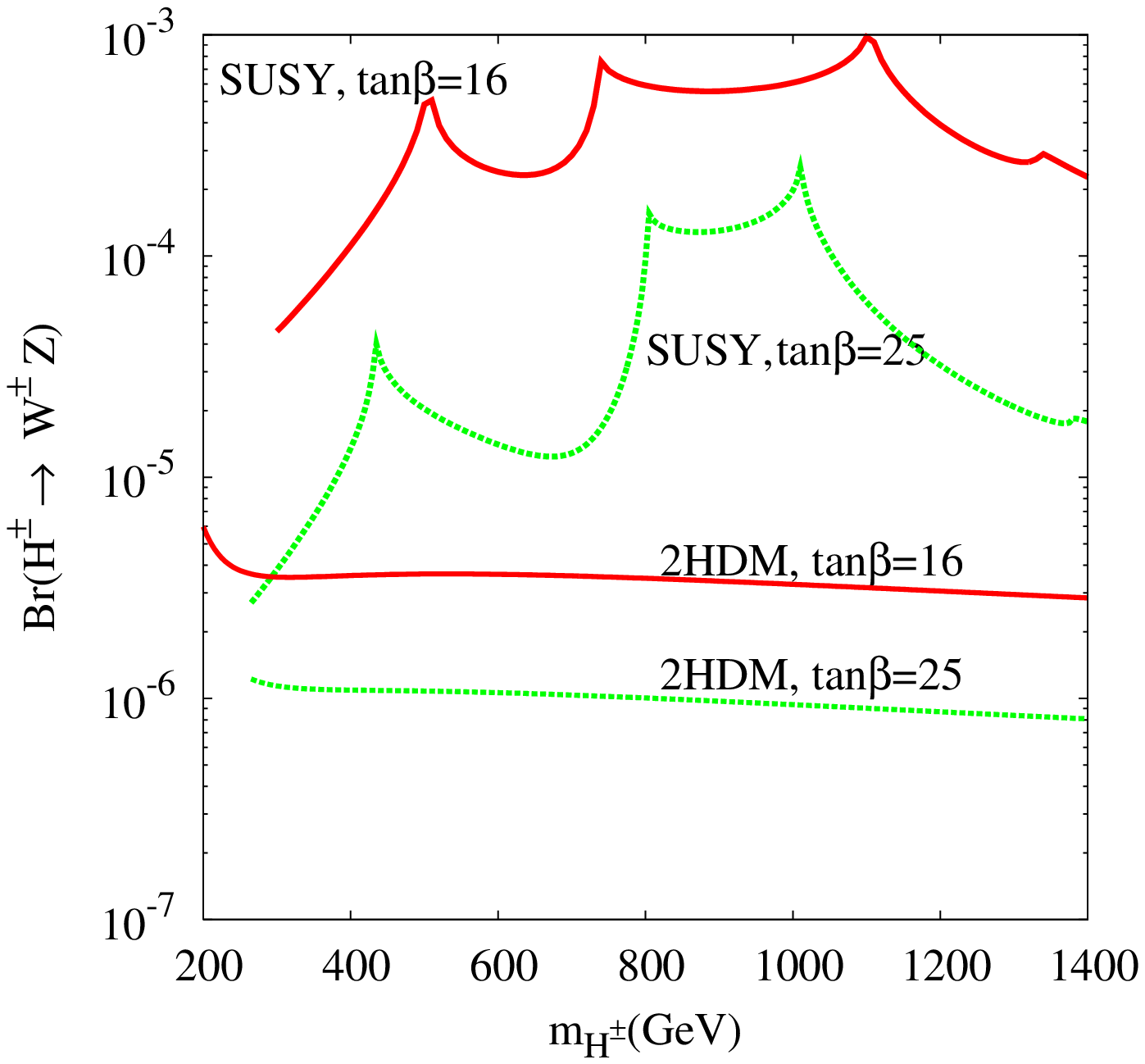}} &
\hspace{-2cm}\resizebox{78mm}{!}{\includegraphics{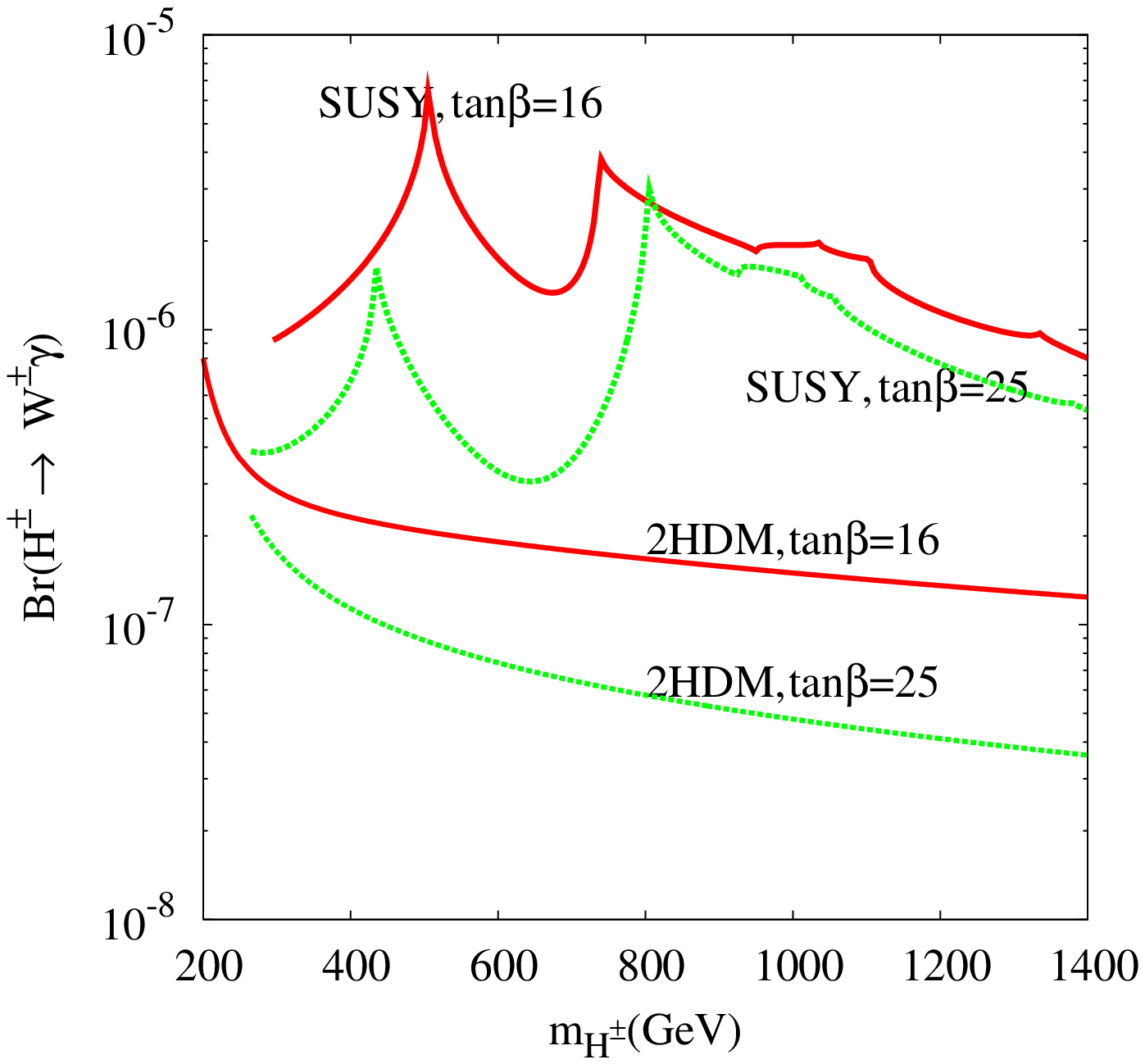}}
\end{tabular}
\end{center}
\caption
{Branching ratios of $H^{\pm}\to W^{\pm}Z$ (left) and 
$H^{\pm}\to W^{\pm}\gamma$ (right) as a function of 
$m_{H^{\pm}}$  in  the MSSM and 2HDM 
for $M_{SUSY}=500$ GeV, $M_2=175$ GeV, $\mu = -1.4$ TeV and
 $A_t =A_b = A_{\tau}= -\mu$ for various values of $\tan\beta$.}
\label{nfig1}
\end{figure}
%=================================
In Fig.~\ref{nfig1}, we show branching ratios of 
$H^{\pm}\to W^{\pm}Z$ (left) and  $H^{\pm}\to W^{\pm}\gamma$ (right)
as a function of charged Higgs mass for $\tan\beta=16$ and 25. 
In those plots, we have shown both the
pure 2HDM \footnote{Pure 2HDM means that we include just the 2HDM part of the
MSSM that contributes here in the loop, {\it i.e.}
only SM fermions, gauge bosons and Higgs bosons with 
MSSM sum rules for the Higgs sector.} and the full MSSM 
contributions. As it can be seen from those plots,
both for $H^\pm \to W^\pm Z$ and $H^\pm\to W^\pm\gamma$ the 
2HDM contribution is rather small. Once we include the SUSY particles, 
we can see that the branching fractions get enhanced and 
can reach $10^{-3}$ in the case
of $H^\pm \to W^\pm Z$ and $10^{-5}$ in the case of $H^\pm \to W^\pm \gamma$. 
The source of this enhancement is mainly due 
to the presence of scalar fermions contributions in the loop which are
amplified by threshold effects from the opening of the decay 
$H^\pm \to \widetilde{t}_i\widetilde{b}_j^*$. It turns out that the 
contribution of charginos neutralinos loops do not enhance the 
branching fractions significantly as compared to scalar fermions loops.
The plots also show that, the branching fractions are more important for 
intermediate $\tan\beta=16$ and are slightly reduced for larger $\tan\beta=25$.

%=================================
\begin{figure}
\begin{tabular}{cc}
\resizebox{78mm}{!}{\includegraphics{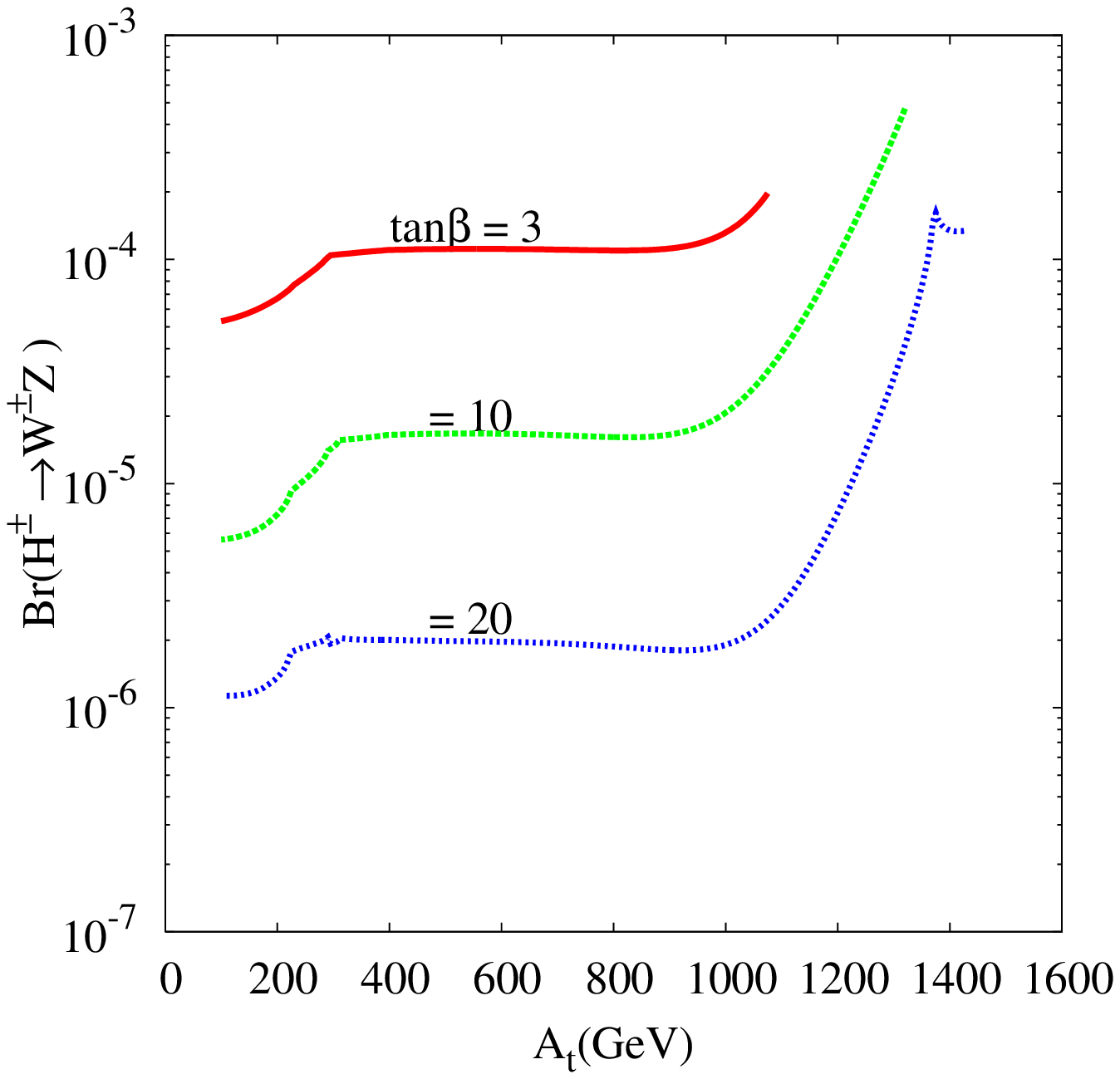}} &
\hspace{-2cm}\resizebox{78mm}{!}{\includegraphics{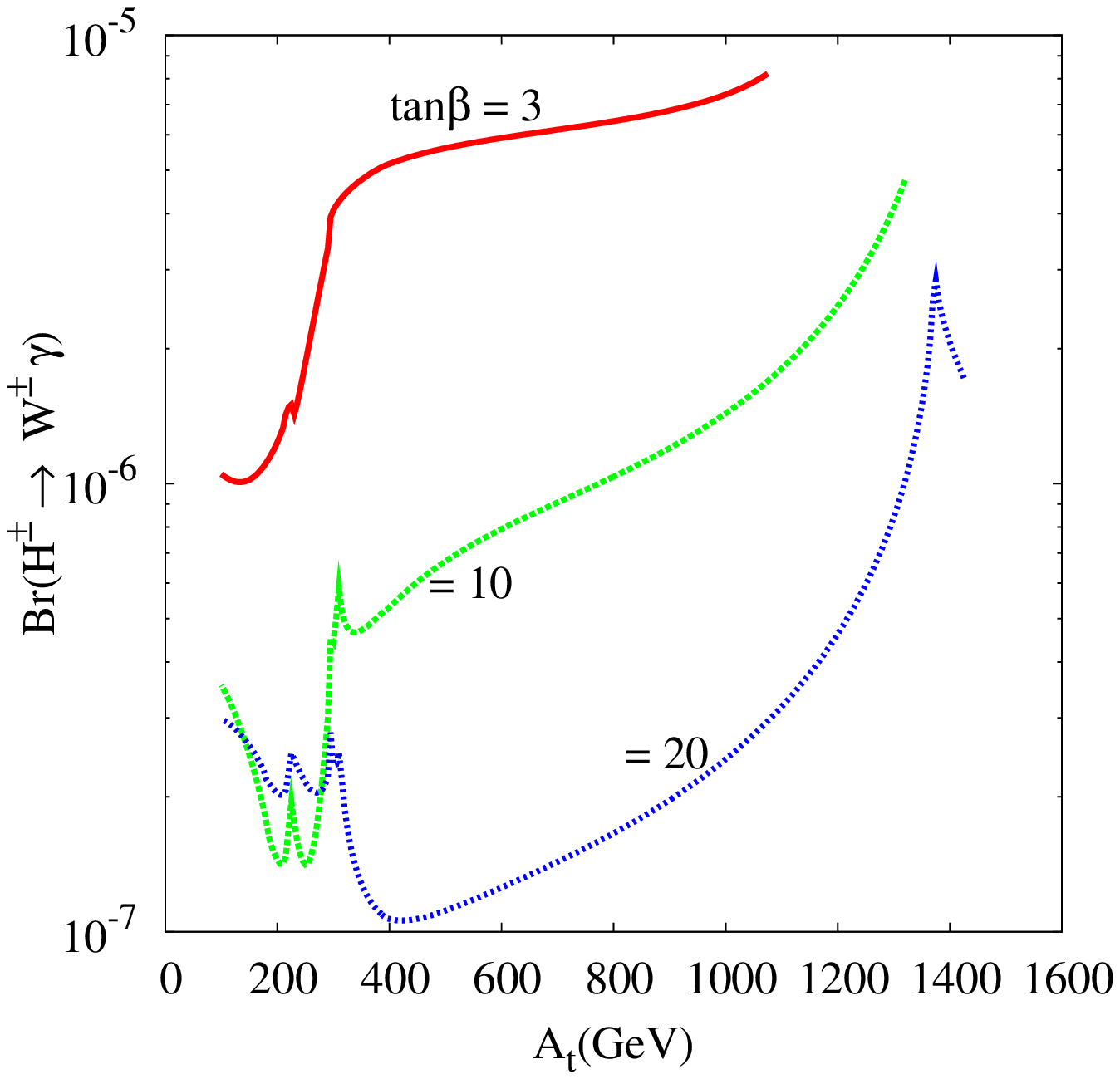}}
\end{tabular}
\caption
{Branching ratios for $H^{\pm}\to W^{\pm}Z$ (left) and
$H^{\pm}\to W^{\pm}\gamma$ (right) as a function of $A_t$  
in the MSSM with $M_{SUSY}=500$ GeV, $M_2=200$ GeV, 
$m_{H^{\pm}}=500$ GeV, $A_t = A_b= A_{\tau}= -\mu$ and $-2\, 
\rm{TeV} < \mu < -0.1$ TeV for various values of $\tan\beta$. }
\label{nfig3}
\end{figure}
%%%%%%%%%%%%%%%%%%%%%%%%%%%%%%%%%%%%%%%%%%%%%%%%%%%%%%%%%%%%%

\begin{figure}
\begin{tabular}{cc}
\hspace{-1.7cm}
\resizebox{79mm}{!}{\includegraphics{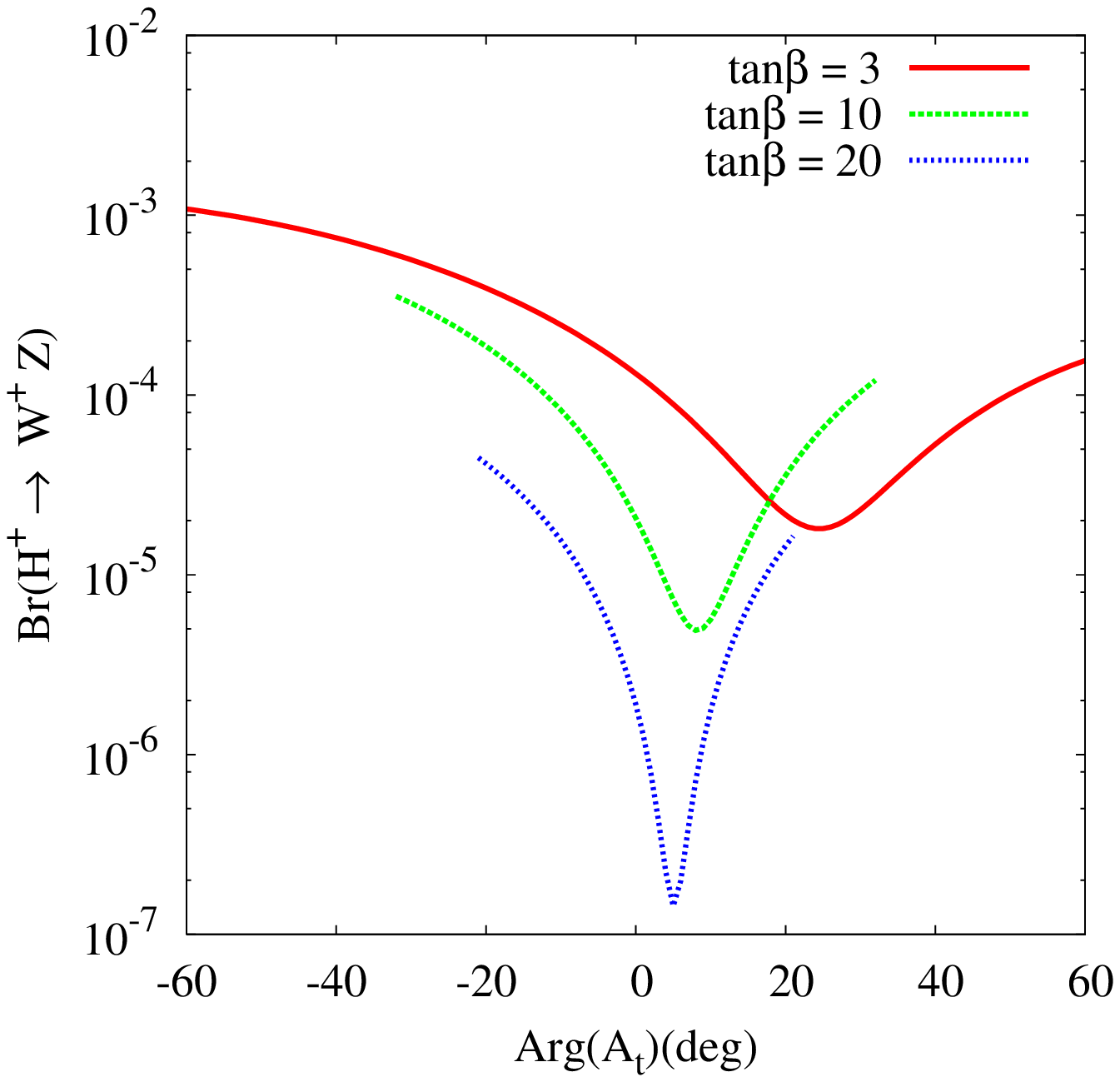}} &
\hspace{-2cm}\resizebox{79mm}{!}{\includegraphics{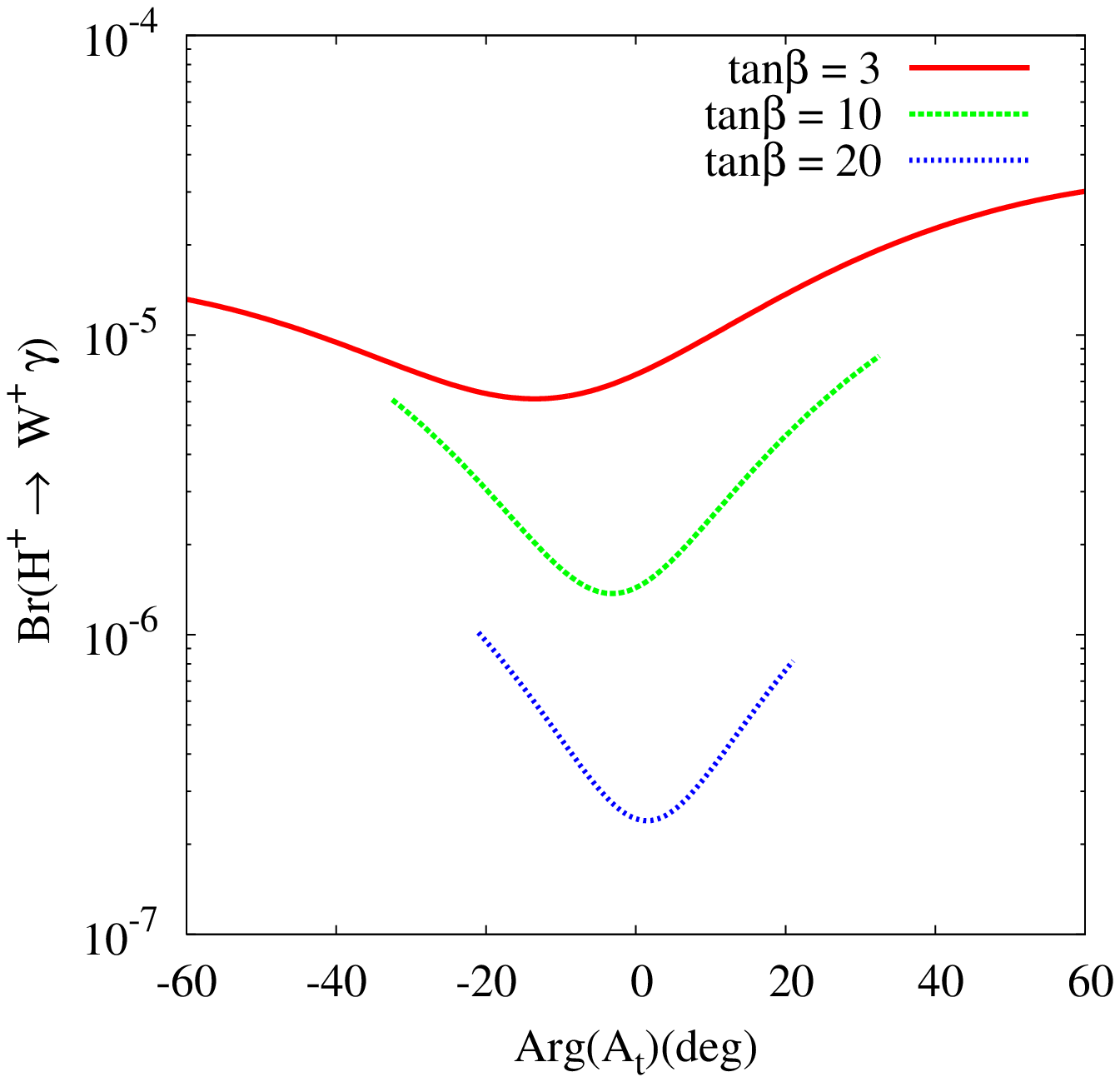}}
\end{tabular}
\caption
{Branching ratios for $H^{\pm}\to W^{\pm}Z$ (left) 
and $H^{\pm}\to W^{\pm}\gamma$ (right) in the
 MSSM as a function of $Arg(A_t)$ :
$M_{SUSY}=500$ GeV, $M_2=150\,GeV$, $m_{H^{\pm}}=500$ GeV, 
$A_t=A_b=A_{\tau}=-\mu =1$ TeV and for various values of $\tan\beta$.}
\label{figcp}
\end{figure}

We now illustrate in Fig.~\ref{nfig3} the branching fraction of 
$H^\pm\to W^\pm Z$ (left) and $H^\pm \to W^\pm \gamma$ (right) 
as a function of $A_t=A_b=A_{\tau}=-\mu $ for $M_{SUSY}=500$ GeV
and $M_2=200$ GeV. 
Since $b\to s \gamma$ favors $A_t$ and $\mu$ to have opposite sign,
we fix $\mu=-A_t$ and in this sense $\mu$ is also varied when 
$A_t$ is varied. Both for $H^\pm\to W^\pm Z$ and $H^\pm\to W^\pm \gamma$,
the chargino-neutralino contributions, which are rather small,
decrease with $\mu=-A_t$: the larger $A_t$, the smaller  
chargino-neutralino contributions.
As one can see from these figures, the plots stops at 
$A_t=1.1$ TeV and $\tan\beta=3$, because for larger $A_t$ the $\delta\rho$ constraint 
will be violated. For $\tan\beta=10$ and 20,
the plots stop for the same reason.
In case of $H^\pm\to W^\pm Z$, for $A_t\la 1$ TeV it is 
the pure 2HDM contributions which dominate and that is why it is 
almost independent of $A_t$, while for larger $A_t$ 
the branching ratios increase with $A_t$. 
It is clear that the larger $A_t$ is the 
larger the branching ratios, which can be of the order of $10^{-3}$ for
$H^\pm\to W^\pm Z$ with $\tan\beta=10$, for example. 
As we know from $h^0\to \gamma \gamma$ and 
$h^0\to \gamma Z$ in MSSM \cite{whk},
the squarks contributions decouple except in the light stop mass 
and large $A_t$ limit \cite{whk}. 
In $H^\pm \to W^\pm V$ case, the same situation happens.
As we can see from Fig.~\ref{nfig3} (left), for intermediate $A_t$,
$300<A_t<1000$ GeV, the squarks are rather heavy and hence their contributions
are small compared to 2HDM one. While for large $A_t$ 
the stop becomes very light $\la 200$ GeV and hence it can enhance the
$H^\pm \to W^\pm V$ width.
Of course this enhancement is also amplified by
$H^\pm\widetilde{t}_{L,R}\widetilde{b}_{R,L}^*$ 
 couplings which are directly proportional to $A_{t,b}$. 
In the case of $H^\pm\to W^\pm \gamma$ decay,
the pure 2HDM and sfermions contribution are of comparable size,
the branching ratio increases with $A_t$. 
We have also studied the effect of the MSSM CP violating phases
on charged Higgs decays. 
For $H^\pm \to W^\pm Z$ and $H^\pm \to W^\pm \gamma$ decays
which are sensitive to MSSM CP violating phases through squarks and 
charginos-neutralinos contributions,
it turns out that the effect of MSSM CP violating phases is important
and can enhance the rate by about one order of
magnitude. For illustration we show in Fig.~\ref{figcp} the effect
of $A_{t,b,\tau} $ CP violating phases for $M_{SUSY}=500$ GeV, 
$A_{t,b,\tau}=-\mu=$ 1 TeV and $M_2=150$ GeV. 
To avoid conflict with EDM constraint we set the $\mu$ parameter
to be real \cite{nath}. As it is clear,
the CP phase of $A_{t,b,\tau}$ can enhance the rates of both 
$H^\pm \to W^\pm (Z, \gamma)$ by more than one order of magnitude.
The observed cuts in the plot are due to $b\to s \gamma$ constraint.

%=================================
%=================================
\begin{figure}
\smallskip\smallskip 
%\vskip2cm
\centerline{{
\epsfxsize3.2 in 
\epsffile{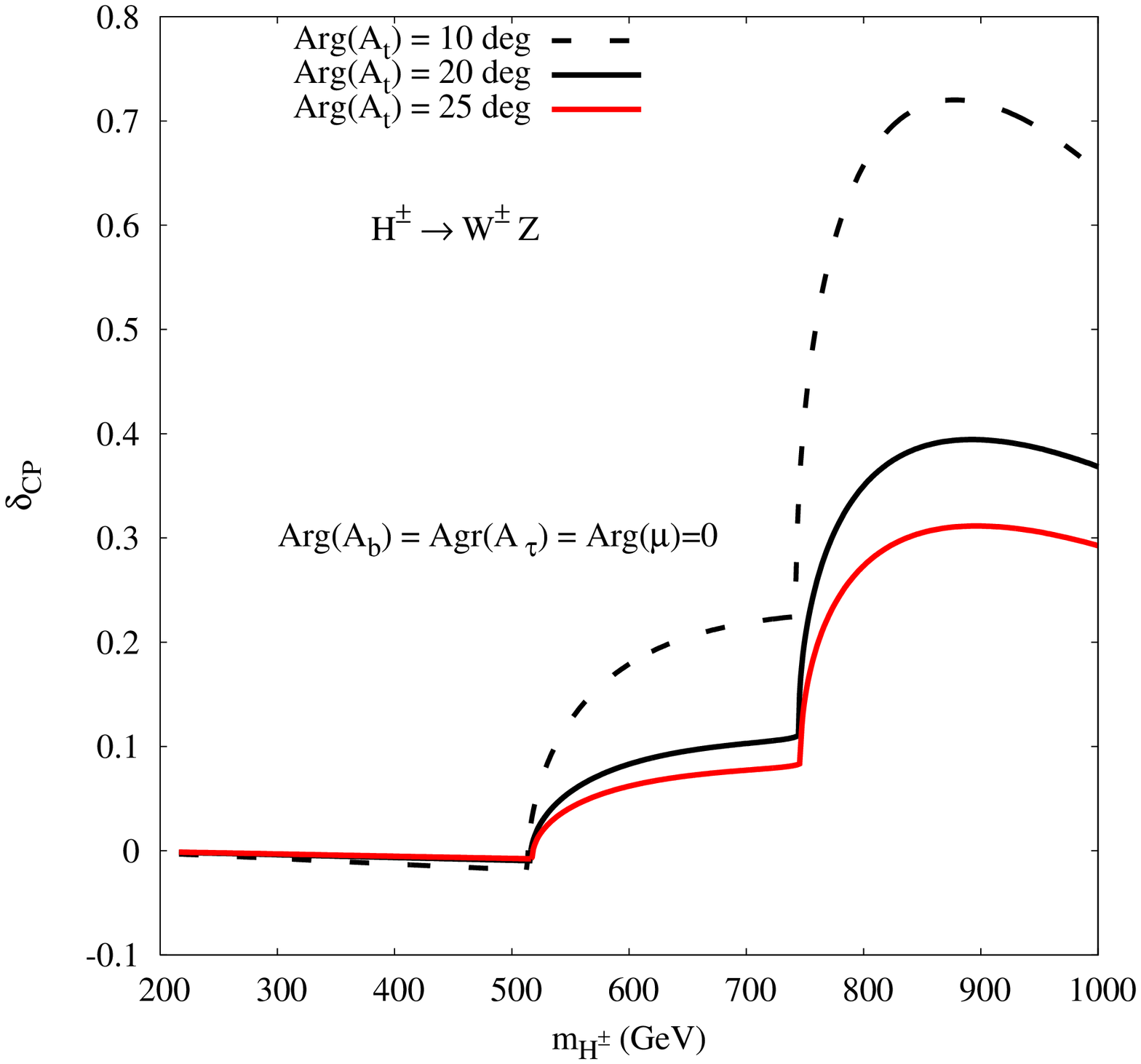}}
\hskip-2cm
\epsfxsize3.2 in 
\epsffile{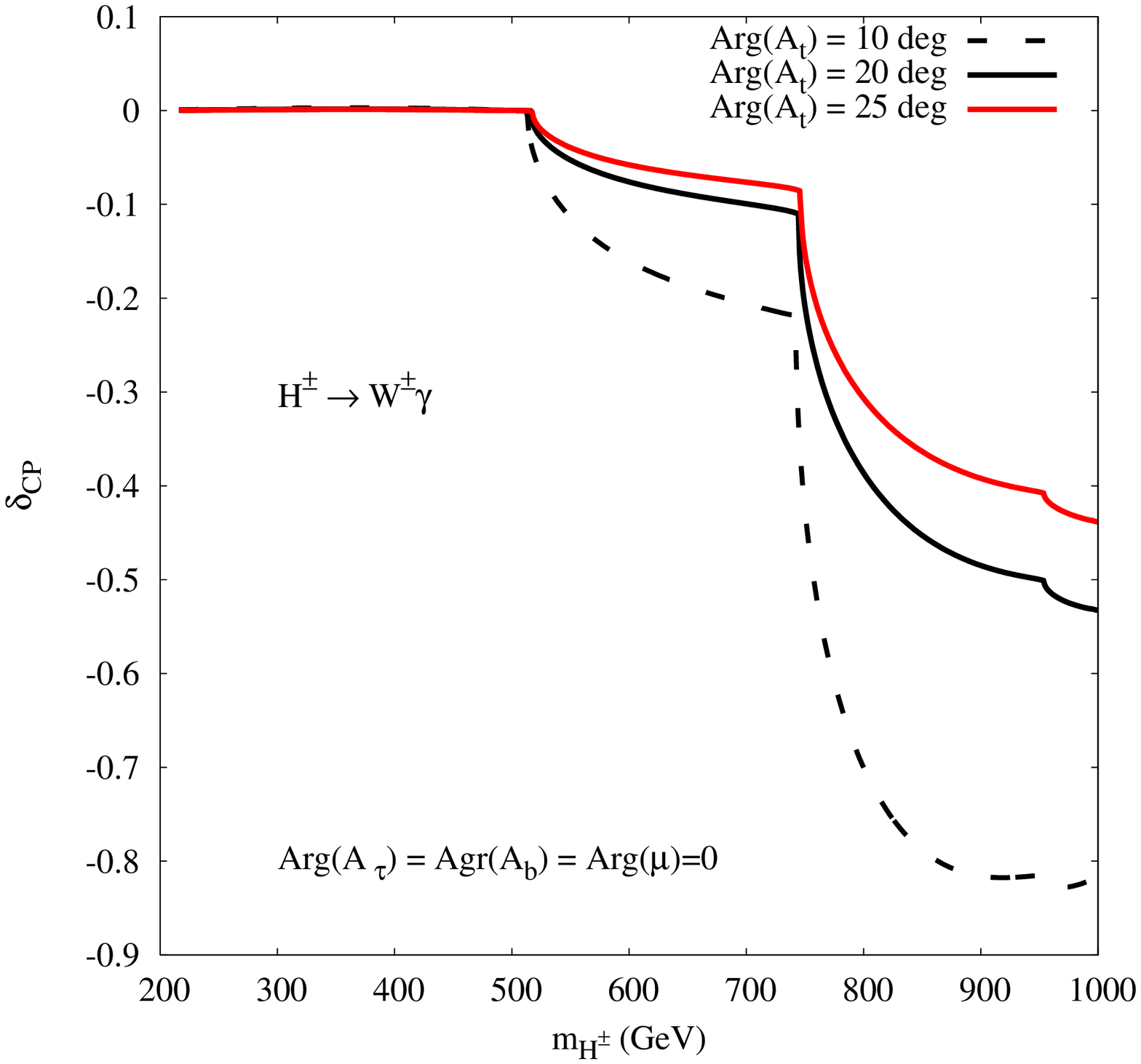}}
\smallskip\smallskip 
\caption{Total $\delta_{CP}$ in the $H^\pm\to W^\pm Z$ (left) and
$H^\pm\to W^\pm \gamma$ (right)  as a function of $m_{H^\pm}$ in the MSSM, 
with $M_{SUSY}=500$ GeV, $M_2= 175$\,GeV, $\tan\beta=16$, $\mu = -1.4$ TeV,
 $A_t=A_b=A_{tau}=-\mu$, for different values of Arg($A_t$).}
\label{cpafig}
\end{figure}

In Fig.(\ref{cpafig}) we show the CP asymmetry $\delta_{CP}$
for $H^\pm\to W^\pm Z$ (left) and $H^\pm\to W^\pm \gamma$ 
(right) as a function of charged Higgs mass. 
As one can see in both plots, for $m_{H\pm}\la 512$ GeV, 
$\delta_{CP}$ is very small, of the order $10^{-3}$. 
In this case, the main contributions to $\delta_{CP}$ arise both from diagrams 
Fig.1.1 and Fig.1.18 with absorptive parts and 
real couplings and diagrams
Fig.1.2 and Fig.1.19 with complex couplings and no absorptive part
 (the channel $H^\pm \to \widetilde{t}_i  \widetilde{b}_j^*$ is not yet open!).
However, once $m_{H\pm}\ga 512$ GeV, the channel 
$H^\pm \to \widetilde{t}_i  \widetilde{b}_j^*$ is open, 
the diagrams Fig.1.2 and Fig.1.19 have an absorptive part and consequently they
contribute to $\delta_{CP}$, which can go up to several percent.
The thresholds of  $H^\pm \to \widetilde{t}_1  \widetilde{b}_1^*$
at $m_{H\pm}\approx 512$ GeV and 
of $H^\pm \to \widetilde{t}_1 \widetilde{b}_2^*$ at $m_{H\pm}\approx 740$ GeV 
are clearly visible in Fig.(\ref{cpafig}).

\section{Conclusion}
In the framework of MSSM, we have studied
charged Higgs decays into a pair of gauge bosons, namely,
$H^\pm\to W^\pm Z$ and $H^\pm\to W^\pm \gamma $.
We have also studied the effects of MSSM CP violating 
phases in these processes.
In contrast to previous studies, we have performed the calculation in
the 't Hooft-Feynman gauge and used a renormalization  prescription to deal
with tadpoles, $W^\pm$--$H^\pm$ and $G^\pm$--$H^\pm$ mixings.
The study has been carried out taking into account the experimental 
constraints on the $\rho$ parameter as well as the inclusive $b$-decay
$b\to s \gamma$.
Numerical results for the branching ratios have been presented.
In the MSSM, we have shown that the  branching ratio
of $H^\pm \to W^\pm Z$ can reach $10^{-3}$ in some cases 
while $H^\pm \to W^\pm \gamma$ never exceed $10^{-5}$.
Branching ratio of the order $10^{-3}$ might provide 
an opportunity to search for a charged Higgs boson at the LHC 
through $H^\pm \to W^{\pm} Z$.
The effect of MSSM CP violating phases is also found to be important
 and can change the size of the branching ratios of  $H^\pm \to W^\pm V$
by about an order of magnitude compared to the CP conserving case.
The CP-violating asymmetry in the decays $H^+ \to W^+V$ and 
$H^- \to W^-V$ can be rather important and can reach 80\% in some
region of parameter space [\refcite{prep}].

\section*{Acknowledgments} 
This work is partially supported by the ICTP through the OEA-project-49
and  by the Moroccan program PROTARS-III D16/04. 
AA, RB, WTC and TCY are supported by the National Science Council of
R.O.C. under Grant Nos. NSC96-2811-M-008-020 , 
NSC96-2811-M-033-002 and NSC 95-2112-M-007-001.

\end{document}

%% file: vertz_MSSM.tex
\unitlength=1bp%

\begin{feynartspicture}(372,504)(5,5.3)
%\FALabel(55.,113.96)[]{\large $H\quad \to\quad W\quad Z$}

\FADiagram{1.1a}
\FAProp(0.,10.)(6.5,10.)(0.,){/ScalarDash}{1}
\FALabel(3.25,8.93)[t]{$H$}
\FAProp(20.,15.)(13.,14.)(0.,){/Sine}{-1}
\FALabel(16.2808,15.5544)[b]{$W$}
\FAProp(20.,5.)(13.,6.)(0.,){/Sine}{0}
\FALabel(16.2808,4.44558)[t]{$V$}
\FAProp(6.5,10.)(13.,14.)(0.,){/Straight}{1}
\FALabel(9.20801,13.1807)[br]{$t$}
\FAProp(6.5,10.)(13.,6.)(0.,){/Straight}{-1}
\FALabel(9.20801,6.81927)[tr]{$b$}
\FAProp(13.,14.)(13.,6.)(0.,){/Straight}{1}
\FALabel(14.274,10.)[l]{$b$}
\FAVert(6.5,10.){0}
\FAVert(13.,14.){0}
\FAVert(13.,6.){0}

\FADiagram{1.1b}
\FAProp(0.,10.)(6.5,10.)(0.,){/ScalarDash}{1}
\FALabel(3.25,8.93)[t]{$H$}
\FAProp(20.,15.)(13.,14.)(0.,){/Sine}{-1}
\FALabel(16.2808,15.5544)[b]{$W$}
\FAProp(20.,5.)(13.,6.)(0.,){/Sine}{0}
\FALabel(16.2808,4.44558)[t]{$V$}
\FAProp(6.5,10.)(13.,14.)(0.,){/Straight}{-1}
\FALabel(9.20801,13.1807)[br]{$b$}
\FAProp(6.5,10.)(13.,6.)(0.,){/Straight}{1}
\FALabel(9.20801,6.81927)[tr]{$t$}
\FAProp(13.,14.)(13.,6.)(0.,){/Straight}{-1}
\FALabel(14.274,10.)[l]{$t$}
\FAVert(6.5,10.){0}
\FAVert(13.,14.){0}
\FAVert(13.,6.){0}

\FADiagram{1.1c}
\FAProp(0.,10.)(6.5,10.)(0.,){/ScalarDash}{1}
\FALabel(3.25,8.93)[t]{$H$}
\FAProp(20.,15.)(13.,14.)(0.,){/Sine}{-1}
\FALabel(16.2808,15.5544)[b]{$W$}
\FAProp(20.,5.)(13.,6.)(0.,){/Sine}{0}
\FALabel(16.2808,4.44558)[t]{$V$}
\FAProp(6.5,10.)(13.,14.)(0.,){/Straight}{1}
\FALabel(9.20801,13.1807)[br]{$\tilde\chi_i$}
\FAProp(6.5,10.)(13.,6.)(0.,){/Straight}{0}
\FALabel(9.33903,7.03218)[tr]{$\tilde \chi_j$}
\FAProp(13.,14.)(13.,6.)(0.,){/Straight}{0}
\FALabel(14.024,10.)[l]{$\tilde \chi_k$}
\FAVert(6.5,10.){0}
\FAVert(13.,14.){0}
\FAVert(13.,6.){0}

\FADiagram{1.2}
\FAProp(0.,10.)(6.5,10.)(0.,){/ScalarDash}{1}
\FALabel(3.25,8.93)[t]{$H$}
\FAProp(20.,15.)(13.,14.)(0.,){/Sine}{-1}
\FALabel(16.2808,15.5544)[b]{$W$}
\FAProp(20.,5.)(13.,6.)(0.,){/Sine}{0}
\FALabel(16.2808,4.44558)[t]{$V$}
\FAProp(6.5,10.)(13.,14.)(0.,){/ScalarDash}{0}
\FALabel(9.33903,12.9678)[br]{$S_i$}
\FAProp(6.5,10.)(13.,6.)(0.,){/ScalarDash}{0}
\FALabel(9.33903,7.03218)[tr]{$S_j$}
\FAProp(13.,14.)(13.,6.)(0.,){/ScalarDash}{0}
\FALabel(14.024,10.)[l]{$S_k$}
\FAVert(6.5,10.){0}
\FAVert(13.,14.){0}
\FAVert(13.,6.){0}

\FADiagram{1.3}
\FAProp(0.,10.)(6.5,10.)(0.,){/ScalarDash}{1}
\FALabel(3.25,8.93)[t]{$H$}
\FAProp(20.,15.)(13.,14.)(0.,){/Sine}{-1}
\FALabel(16.2808,15.5544)[b]{$W$}
\FAProp(20.,5.)(13.,6.)(0.,){/Sine}{0}
\FALabel(16.2808,4.44558)[t]{$V$}
\FAProp(6.5,10.)(13.,14.)(0.,){/ScalarDash}{0}
\FALabel(9.33903,12.9678)[br]{$S_i$}
\FAProp(6.5,10.)(13.,6.)(0.,){/ScalarDash}{0}
\FALabel(9.33903,7.03218)[tr]{$S_j$}
\FAProp(13.,14.)(13.,6.)(0.,){/Sine}{0}
\FALabel(14.274,10.)[l]{$V$}
\FAVert(6.5,10.){0}
\FAVert(13.,14.){0}
\FAVert(13.,6.){0}

\FADiagram{1.4}
\FAProp(0.,10.)(6.5,10.)(0.,){/ScalarDash}{1}
\FALabel(3.25,8.93)[t]{$H$}
\FAProp(20.,15.)(13.,14.)(0.,){/Sine}{-1}
\FALabel(16.2808,15.5544)[b]{$W$}
\FAProp(20.,5.)(13.,6.)(0.,){/Sine}{0}
\FALabel(16.2808,4.44558)[t]{$Z$}
\FAProp(6.5,10.)(13.,14.)(0.,){/Sine}{0}
\FALabel(9.20801,13.1807)[br]{$V$}
\FAProp(6.5,10.)(13.,6.)(0.,){/ScalarDash}{0}
\FALabel(9.33903,7.03218)[tr]{$S_i$}
\FAProp(13.,14.)(13.,6.)(0.,){/ScalarDash}{0}
\FALabel(14.024,10.)[l]{$S_j$}
\FAVert(6.5,10.){0}
\FAVert(13.,14.){0}
\FAVert(13.,6.){0}

\FADiagram{1.5}
\FAProp(0.,10.)(6.5,10.)(0.,){/ScalarDash}{1}
\FALabel(3.25,8.93)[t]{$H$}
\FAProp(20.,15.)(13.,14.)(0.,){/Sine}{-1}
\FALabel(16.2808,15.5544)[b]{$W$}
\FAProp(20.,5.)(13.,6.)(0.,){/Sine}{0}
\FALabel(16.2808,4.44558)[t]{$V$}
\FAProp(6.5,10.)(13.,14.)(0.,){/ScalarDash}{0}
\FALabel(9.33903,12.9678)[br]{$S_i$}
\FAProp(6.5,10.)(13.,6.)(0.,){/Sine}{0}
\FALabel(9.20801,6.81927)[tr]{$V$}
\FAProp(13.,14.)(13.,6.)(0.,){/ScalarDash}{0}
\FALabel(14.024,10.)[l]{$S_j$}
\FAVert(6.5,10.){0}
\FAVert(13.,14.){0}
\FAVert(13.,6.){0}

\FADiagram{1.6}
\FAProp(0.,10.)(6.5,10.)(0.,){/ScalarDash}{1}
\FALabel(3.25,8.93)[t]{$H$}
\FAProp(20.,15.)(13.,14.)(0.,){/Sine}{-1}
\FALabel(16.2808,15.5544)[b]{$W$}
\FAProp(20.,5.)(13.,6.)(0.,){/Sine}{0}
\FALabel(16.2808,4.44558)[t]{$V$}
\FAProp(6.5,10.)(13.,14.)(0.,){/ScalarDash}{0}
\FALabel(9.33903,12.9678)[br]{$S_i$}
\FAProp(6.5,10.)(13.,6.)(0.,){/Sine}{0}
\FALabel(9.20801,6.81927)[tr]{$V$}
\FAProp(13.,14.)(13.,6.)(0.,){/Sine}{0}
\FALabel(14.274,10.)[l]{$V$}
\FAVert(6.5,10.){0}
\FAVert(13.,14.){0}
\FAVert(13.,6.){0}

\FADiagram{1.7}
\FAProp(0.,10.)(6.5,10.)(0.,){/ScalarDash}{1}
\FALabel(3.25,8.93)[t]{$H$}
\FAProp(20.,15.)(13.,14.)(0.,){/Sine}{-1}
\FALabel(16.2808,15.5544)[b]{$W$}
\FAProp(20.,5.)(13.,6.)(0.,){/Sine}{0}
\FALabel(16.2808,4.44558)[t]{$V$}
\FAProp(6.5,10.)(13.,14.)(0.,){/Sine}{0}
\FALabel(9.20801,13.1807)[br]{$V$}
\FAProp(6.5,10.)(13.,6.)(0.,){/ScalarDash}{0}
\FALabel(9.33903,7.03218)[tr]{$S_i$}
\FAProp(13.,14.)(13.,6.)(0.,){/Sine}{0}
\FALabel(14.274,10.)[l]{$V$}
\FAVert(6.5,10.){0}
\FAVert(13.,14.){0}
\FAVert(13.,6.){0}

\FADiagram{1.8}
\FAProp(0.,10.)(7.,10.)(0.,){/ScalarDash}{1}
\FALabel(3.5,8.93)[t]{$H$}
\FAProp(20.,15.)(13.,10.)(0.,){/Sine}{-1}
\FALabel(16.0791,13.2813)[br]{$W$}
\FAProp(20.,5.)(13.,10.)(0.,){/Sine}{0}
\FALabel(16.9209,8.28129)[bl]{$V$}
\FAProp(7.,10.)(13.,10.)(0.833333,){/ScalarDash}{0}
\FALabel(10.,6.68)[t]{$S_i$}
\FAProp(7.,10.)(13.,10.)(-0.833333,){/ScalarDash}{0}
\FALabel(10.,13.32)[b]{$S_j$}
\FAVert(7.,10.){0}
\FAVert(13.,10.){0}

\FADiagram{1.9}
\FAProp(0.,10.)(6.,10.)(0.,){/ScalarDash}{1}
\FALabel(3.,8.93)[t]{$H$}
\FAProp(20.,15.)(11.5,10.)(0.,){/Sine}{-1}
\FALabel(15.4441,13.356)[br]{$W$}
\FAProp(20.,5.)(11.5,10.)(0.,){/Sine}{0}
\FALabel(16.0559,8.356)[bl]{$V$}
\FAProp(11.5,10.)(6.,10.)(0.,){/ScalarDash}{0}
\FALabel(8.75,9.18)[t]{$S_i$}
\FAProp(6.,10.)(6.,10.)(6.,15.){/ScalarDash}{0}
\FALabel(6.,15.82)[b]{$S_j$}
\FAVert(11.5,10.){0}
\FAVert(6.,10.){0}

\FADiagram{1.10}
\FAProp(0.,10.)(10.5,8.)(0.,){/ScalarDash}{1}
\FALabel(4.95998,7.95738)[t]{$H$}
\FAProp(20.,15.)(12.5,13.5)(0.,){/Sine}{-1}
\FALabel(15.946,15.2899)[b]{$W$}
\FAProp(20.,5.)(10.5,8.)(0.,){/Sine}{0}
\FALabel(14.7832,5.50195)[t]{$V$}
\FAProp(12.5,13.5)(10.5,8.)(0.8,){/ScalarDash}{0}
\FALabel(8.55827,11.9943)[r]{$S_i$}
\FAProp(12.5,13.5)(10.5,8.)(-0.8,){/Sine}{0}
\FALabel(14.6767,9.4203)[l]{$V$}
\FAVert(12.5,13.5){0}
\FAVert(10.5,8.){0}

\FADiagram{1.11}
\FAProp(0.,10.)(9.5,12.)(0.,){/ScalarDash}{1}
\FALabel(4.43068,12.0368)[b]{$H$}
\FAProp(20.,15.)(9.5,12.)(0.,){/Sine}{-1}
\FALabel(14.3242,14.5104)[b]{$W$}
\FAProp(20.,5.)(12.5,6.5)(0.,){/Sine}{0}
\FALabel(15.946,4.7101)[t]{$V$}
\FAProp(12.5,6.5)(9.5,12.)(0.8,){/ScalarDash}{0}
\FALabel(14.12,10.22)[bl]{$S_i$}
\FAProp(12.5,6.5)(9.5,12.)(-0.8,){/Sine}{0}
\FALabel(7.91926,7.78778)[tr]{$V$}
\FAVert(12.5,6.5){0}
\FAVert(9.5,12.){0}

\FADiagram{1.12}
\FAProp(0.,10.)(11.,10.)(0.,){/ScalarDash}{1}
\FALabel(5.5,8.93)[t]{$H$}
\FAProp(20.,15.)(17.3,13.5)(0.,){/Sine}{-1}
\FALabel(18.3773,15.1249)[br]{$W$}
\FAProp(20.,5.)(11.,10.)(0.,){/Sine}{0}
\FALabel(15.2273,6.62506)[tr]{$V$}
\FAProp(11.,10.)(13.7,11.5)(0.,){/ScalarDash}{0}
\FALabel(12.1987,11.4064)[br]{$S_i$}
\FAProp(17.3,13.5)(13.7,11.5)(-0.8,){/Straight}{0}
\FALabel(16.4513,10.4036)[tl]{$F_j$}
\FAProp(17.3,13.5)(13.7,11.5)(0.8,){/Straight}{0}
\FALabel(14.5487,14.5964)[br]{$F_k$}
\FAVert(11.,10.){0}
\FAVert(17.3,13.5){0}
\FAVert(13.7,11.5){0}

\FADiagram{1.13}
\FAProp(0.,10.)(11.,10.)(0.,){/ScalarDash}{1}
\FALabel(5.5,8.93)[t]{$H$}
\FAProp(20.,15.)(17.3,13.5)(0.,){/Sine}{-1}
\FALabel(18.3773,15.1249)[br]{$W$}
\FAProp(20.,5.)(11.,10.)(0.,){/Sine}{0}
\FALabel(15.2273,6.62506)[tr]{$V$}
\FAProp(11.,10.)(13.7,11.5)(0.,){/ScalarDash}{0}
\FALabel(12.1987,11.4064)[br]{$S_i$}
\FAProp(17.3,13.5)(13.7,11.5)(-0.8,){/ScalarDash}{0}
\FALabel(16.4513,10.4036)[tl]{$S_j$}
\FAProp(17.3,13.5)(13.7,11.5)(0.8,){/ScalarDash}{0}
\FALabel(14.5487,14.5964)[br]{$S_k$}
\FAVert(11.,10.){0}
\FAVert(17.3,13.5){0}
\FAVert(13.7,11.5){0}

\FADiagram{1.14}
\FAProp(0.,10.)(11.,10.)(0.,){/ScalarDash}{1}
\FALabel(5.5,8.93)[t]{$H$}
\FAProp(20.,15.)(17.3,13.5)(0.,){/Sine}{-1}
\FALabel(18.3773,15.1249)[br]{$W$}
\FAProp(20.,5.)(11.,10.)(0.,){/Sine}{0}
\FALabel(15.2273,6.62506)[tr]{$V$}
\FAProp(11.,10.)(13.7,11.5)(0.,){/ScalarDash}{0}
\FALabel(12.1987,11.4064)[br]{$S_i$}
\FAProp(17.3,13.5)(13.7,11.5)(-0.8,){/ScalarDash}{0}
\FALabel(16.4513,10.4036)[tl]{$S_j$}
\FAProp(17.3,13.5)(13.7,11.5)(0.8,){/Sine}{0}
\FALabel(14.4273,14.8149)[br]{$V$}
\FAVert(11.,10.){0}
\FAVert(17.3,13.5){0}
\FAVert(13.7,11.5){0}

\FADiagram{1.15}
\FAProp(0.,10.)(3.5,10.)(0.,){/ScalarDash}{1}
\FALabel(1.75,8.93)[t]{$H$}
\FAProp(20.,15.)(12.5,10.)(0.,){/Sine}{-1}
\FALabel(15.8702,13.3097)[br]{$W$}
\FAProp(20.,5.)(12.5,10.)(0.,){/Sine}{0}
\FALabel(16.6298,8.30968)[bl]{$V$}
\FAProp(12.5,10.)(9.,10.)(0.,){/ScalarDash}{0}
\FALabel(10.75,9.18)[t]{$S_i$}
\FAProp(3.5,10.)(9.,10.)(0.8,){/Straight}{0}
\FALabel(6.25,6.98)[t]{$F_j$}
\FAProp(3.5,10.)(9.,10.)(-0.8,){/Straight}{0}
\FALabel(6.25,13.02)[b]{$F_k$}
\FAVert(3.5,10.){0}
\FAVert(12.5,10.){0}
\FAVert(9.,10.){0}

\FADiagram{1.16}
\FAProp(0.,10.)(3.5,10.)(0.,){/ScalarDash}{1}
\FALabel(1.75,8.93)[t]{$H$}
\FAProp(20.,15.)(12.5,10.)(0.,){/Sine}{-1}
\FALabel(15.8702,13.3097)[br]{$W$}
\FAProp(20.,5.)(12.5,10.)(0.,){/Sine}{0}
\FALabel(16.6298,8.30968)[bl]{$V$}
\FAProp(12.5,10.)(9.,10.)(0.,){/ScalarDash}{0}
\FALabel(10.75,9.18)[t]{$S_i$}
\FAProp(3.5,10.)(9.,10.)(0.8,){/ScalarDash}{0}
\FALabel(6.25,6.98)[t]{$S_j$}
\FAProp(3.5,10.)(9.,10.)(-0.8,){/ScalarDash}{0}
\FALabel(6.25,13.02)[b]{$S_k$}
\FAVert(3.5,10.){0}
\FAVert(12.5,10.){0}
\FAVert(9.,10.){0}

\FADiagram{1.17}
\FAProp(0.,10.)(3.5,10.)(0.,){/ScalarDash}{1}
\FALabel(1.75,8.93)[t]{$H$}
\FAProp(20.,15.)(12.5,10.)(0.,){/Sine}{-1}
\FALabel(15.8702,13.3097)[br]{$W$}
\FAProp(20.,5.)(12.5,10.)(0.,){/Sine}{0}
\FALabel(16.6298,8.30968)[bl]{$V$}
\FAProp(12.5,10.)(9.,10.)(0.,){/ScalarDash}{0}
\FALabel(10.75,9.18)[t]{$S_i$}
\FAProp(3.5,10.)(9.,10.)(0.8,){/ScalarDash}{0}
\FALabel(6.25,6.98)[t]{$S_j$}
\FAProp(3.5,10.)(9.,10.)(-0.8,){/Sine}{0}
\FALabel(6.25,13.27)[b]{$V$}
\FAVert(3.5,10.){0}
\FAVert(12.5,10.){0}
\FAVert(9.,10.){0}

\FADiagram{1.18}
\FAProp(0.,10.)(3.5,10.)(0.,){/ScalarDash}{1}
\FALabel(1.75,8.93)[t]{$H$}
\FAProp(20.,15.)(12.5,10.)(0.,){/Sine}{-1}
\FALabel(15.8702,13.3097)[br]{$W$}
\FAProp(20.,5.)(12.5,10.)(0.,){/Sine}{0}
\FALabel(16.6298,8.30968)[bl]{$V$}
\FAProp(12.5,10.)(9.,10.)(0.,){/Sine}{0}
\FALabel(10.75,8.93)[t]{$V$}
\FAProp(3.5,10.)(9.,10.)(0.8,){/Straight}{0}
\FALabel(6.25,6.98)[t]{$F_i$}
\FAProp(3.5,10.)(9.,10.)(-0.8,){/Straight}{0}
\FALabel(6.25,13.02)[b]{$F_j$}
\FAVert(3.5,10.){0}
\FAVert(12.5,10.){0}
\FAVert(9.,10.){0}

\FADiagram{1.19}
\FAProp(0.,10.)(3.5,10.)(0.,){/ScalarDash}{1}
\FALabel(1.75,8.93)[t]{$H$}
\FAProp(20.,15.)(12.5,10.)(0.,){/Sine}{-1}
\FALabel(15.8702,13.3097)[br]{$W$}
\FAProp(20.,5.)(12.5,10.)(0.,){/Sine}{0}
\FALabel(16.6298,8.30968)[bl]{$V$}
\FAProp(12.5,10.)(9.,10.)(0.,){/Sine}{0}
\FALabel(10.75,8.93)[t]{$V$}
\FAProp(3.5,10.)(9.,10.)(0.8,){/ScalarDash}{0}
\FALabel(6.25,6.98)[t]{$S_i$}
\FAProp(3.5,10.)(9.,10.)(-0.8,){/ScalarDash}{0}
\FALabel(6.25,13.02)[b]{$S_j$}
\FAVert(3.5,10.){0}
\FAVert(12.5,10.){0}
\FAVert(9.,10.){0}

\FADiagram{1.20}
\FAProp(0.,10.)(3.5,10.)(0.,){/ScalarDash}{1}
\FALabel(1.75,8.93)[t]{$H$}
\FAProp(20.,15.)(12.5,10.)(0.,){/Sine}{-1}
\FALabel(15.8702,13.3097)[br]{$W$}
\FAProp(20.,5.)(12.5,10.)(0.,){/Sine}{0}
\FALabel(16.6298,8.30968)[bl]{$V$}
\FAProp(12.5,10.)(9.,10.)(0.,){/Sine}{0}
\FALabel(10.75,8.93)[t]{$V$}
\FAProp(3.5,10.)(9.,10.)(0.8,){/ScalarDash}{0}
\FALabel(6.25,6.98)[t]{$S_i$}
\FAProp(3.5,10.)(9.,10.)(-0.8,){/Sine}{0}
\FALabel(6.25,13.27)[b]{$V$}
\FAVert(3.5,10.){0}
\FAVert(12.5,10.){0}
\FAVert(9.,10.){0}

\end{feynartspicture}

%% file: self.tex
\unitlength=1bp%

\begin{feynartspicture}(402,504)(4,5.3)
%\FALabel(44.,113.96)[]{\large $H\quad \to\quad W$}

\FADiagram{2.1a}
\FAProp(0.,10.)(6.,10.)(0.,){/ScalarDash}{1}
\FALabel(3.,8.93)[t]{$H$}
\FAProp(20.,10.)(14.,10.)(0.,){/Sine}{-1}
\FALabel(17.,11.07)[b]{$W$}
\FAProp(6.,10.)(14.,10.)(0.8,){/Straight}{0}
\FALabel(10.,5.98)[t]{$t,\tilde\chi$}
\FAProp(6.,10.)(14.,10.)(-0.8,){/Straight}{0}
\FALabel(10.,14.02)[b]{$b,\tilde\chi$}
\FAVert(6.,10.){0}
\FAVert(14.,10.){0}

\FADiagram{2.1b}
\FAProp(0.,10.)(6.,10.)(0.,){/ScalarDash}{1}
\FALabel(3.,8.93)[t]{$H$}
\FAProp(20.,10.)(14.,10.)(0.,){/Sine}{-1}
\FALabel(17.,11.07)[b]{$W$}
\FAProp(6.,10.)(14.,10.)(0.8,){/ScalarDash}{0}
\FALabel(10.,5.98)[t]{$\tilde{t},\tilde{\nu}_l$}
\FAProp(6.,10.)(14.,10.)(-0.8,){/ScalarDash}{0}
\FALabel(10.,14.02)[b]{$\tilde{b},\tilde{l}$}
\FAVert(6.,10.){0}
\FAVert(14.,10.){0}

\FADiagram{2.1c}
\FAProp(0.,10.)(6.,10.)(0.,){/ScalarDash}{1}
\FALabel(3.,8.93)[t]{$H$}
\FAProp(20.,10.)(14.,10.)(0.,){/Sine}{-1}
\FALabel(17.,11.07)[b]{$W$}
\FAProp(6.,10.)(14.,10.)(0.8,){/Sine}{0}
\FALabel(10.,5.98)[t]{$V$}
\FAProp(6.,10.)(14.,10.)(-0.8,){/ScalarDash}{0}
\FALabel(10.,14.02)[b]{$S$}
\FAVert(6.,10.){0}
\FAVert(14.,10.){0}

\FADiagram{2.1d}
\FAProp(0.,10.)(6.,10.)(0.,){/ScalarDash}{1}
\FALabel(3.,8.93)[t]{$H$}
\FAProp(20.,10.)(14.,10.)(0.,){/ScalarDash}{-1}
\FALabel(17.,11.07)[b]{$G$}
\FAProp(6.,10.)(14.,10.)(0.8,){/Straight}{0}
\FALabel(10.,5.98)[t]{$t,\tilde\chi$}
\FAProp(6.,10.)(14.,10.)(-0.8,){/Straight}{0}
\FALabel(10.,14.02)[b]{$b,\tilde\chi$}
\FAVert(6.,10.){0}
\FAVert(14.,10.){0}

\FADiagram{2.1e}
\FAProp(0.,10.)(6.,10.)(0.,){/ScalarDash}{1}
\FALabel(3.,8.93)[t]{$H$}
\FAProp(20.,10.)(14.,10.)(0.,){/ScalarDash}{-1}
\FALabel(17.,11.07)[b]{$G$}
\FAProp(6.,10.)(14.,10.)(0.8,){/ScalarDash}{0}
\FALabel(10.,5.98)[t]{$\tilde{t},\tilde{\nu}_l$}
\FAProp(6.,10.)(14.,10.)(-0.8,){/ScalarDash}{0}
\FALabel(10.,14.02)[b]{$\tilde{b},\tilde{l}$}
\FAVert(6.,10.){0}
\FAVert(14.,10.){0}

\FADiagram{2.1f}
\FAProp(0.,10.)(6.,10.)(0.,){/ScalarDash}{1}
\FALabel(3.,8.93)[t]{$H$}
\FAProp(20.,10.)(14.,10.)(0.,){/ScalarDash}{-1}
\FALabel(17.,11.07)[b]{$G,H$}
\FAProp(6.,10.)(14.,10.)(0.8,){/Sine}{0}
\FALabel(10.,5.98)[t]{$V$}
\FAProp(6.,10.)(14.,10.)(-0.8,){/ScalarDash}{0}
\FALabel(10.,14.02)[b]{$S$}
\FAVert(6.,10.){0}
\FAVert(14.,10.){0}

\FADiagram{2.2a}
\FAProp(0.,10.)(11.,10.)(0.,){/ScalarDash}{1}
\FALabel(5.5,8.93)[t]{$H$}
\FAProp(20.,15.)(11.,10.)(0.,){/Sine}{-1}
\FALabel(15.2273,13.3749)[br]{$W$}
\FAProp(20.,5.)(11.,10.)(0.,){/Sine}{0}
\FALabel(15.2273,6.62506)[tr]{$V$}
\FAVert(11.,10.){1}

\FADiagram{2.2b}
\FAProp(0.,10.)(11.,10.)(0.,){/ScalarDash}{1}
\FALabel(5.5,8.93)[t]{$H$}
\FAProp(20.,15.)(15.5,12.5)(0.,){/Sine}{-1}
\FALabel(17.4773,14.6249)[br]{$W$}
\FAProp(20.,5.)(11.,10.)(0.,){/Sine}{0}
\FALabel(15.2273,6.62506)[tr]{$V$}
\FAProp(15.5,12.5)(11.,10.)(0.,){/ScalarDash}{-1}
\FALabel(12.9773,12.1249)[br]{$H$}
\FAVert(11.,10.){0}
\FAVert(15.5,12.5){1}

\FADiagram{2.2c}
\FAProp(0.,10.)(5.5,10.)(0.,){/ScalarDash}{1}
\FALabel(2.75,8.93)[t]{$H$}
\FAProp(20.,15.)(11.,10.)(0.,){/Sine}{-1}
\FALabel(15.2273,13.3749)[br]{$W$}
\FAProp(20.,5.)(11.,10.)(0.,){/Sine}{0}
\FALabel(15.7727,8.37494)[bl]{$V$}
\FAProp(5.5,10.)(11.,10.)(0.,){/ScalarDash}{1}
\FALabel(8.25,8.93)[t]{$G$}
\FAVert(11.,10.){0}
\FAVert(5.5,10.){1}

\FADiagram{2.2d}
\FAProp(0.,10.)(5.5,10.)(0.,){/ScalarDash}{1}
\FALabel(2.75,8.93)[t]{$H$}
\FAProp(20.,15.)(11.,10.)(0.,){/Sine}{-1}
\FALabel(15.2273,13.3749)[br]{$W$}
\FAProp(20.,5.)(11.,10.)(0.,){/Sine}{0}
\FALabel(15.7727,8.37494)[bl]{$V$}
\FAProp(5.5,10.)(11.,10.)(0.,){/Sine}{1}
\FALabel(8.25,8.93)[t]{$W$}
\FAVert(11.,10.){0}
\FAVert(5.5,10.){1}

%
%\FADiagram{}
%
%\FADiagram{}
%
%\FADiagram{}
%
%\FADiagram{}
%
\end{feynartspicture}